**Title**

THEA: Ontology driven analysis of microarray data

**Authors**

Pasquier[*], C.; Institute of Signaling, Developmental Biology and Cancer Research, Laboratory Of Virtual Biology, CNRS UMR 6543 Centre de Biochimie; Nice; 06108 Cedex 02; France.

Girardot, F.; Institute of Signaling, Developmental Biology and Cancer Research, Laboratory Of Virtual Biology, CNRS UMR 6543 Centre de Biochimie; Nice; 06108 Cedex 02; France.

Jevardat de Fombelle, K.; Institute of Signaling, Developmental Biology and Cancer Research, Laboratory Of Virtual Biology, CNRS UMR 6543 Centre de Biochimie; Nice; 06108 Cedex 02; France.

Christen R.; Institute of Signaling, Developmental Biology and Cancer Research, Laboratory Of Virtual Biology, CNRS UMR 6543 Centre de Biochimie; Nice; 06108 Cedex 02; France.

**Running Head**

Ontology driven analysis of microarray data

**ABSTRACT**

**Motivation**: Microarray technology makes it possible to measure thousands of variables and to compare their values under hundreds of conditions. Once microarray data are quantified, normalized and classified, the analysis phase is essentially a manual and subjective task based on visual inspection of classes in the light of the vast amount of information available. Currently, data interpretation clearly constitutes the bottleneck of such analyses and there is an obvious need for tools able to fill the gap between data processed with mathematical methods and existing biological knowledge.

**Results**: THEA (Tools for High-throughput Experiments Analysis) is an integrated information processing system allowing convenient handling of data. It allows to automatically annotate data issued from classification systems with selected biological information coming from a knowledge base and to either to manually search and browse through these annotations or to automatically generate meaningful generalizations according to statistical criteria (data mining).

**Availability:** The software is available on the web site: http://thea.unice.fr/

**Contact**: claude.pasquier@unice.fr

**Supplementary Information**: Supplementary tables as well as files containing the biological data used in this publication can be downloaded from our website: http://bioinfo.unice.fr/publications/thea_article/

---

[*] To whom correspondence should be addressed

## INTRODUCTION

During the last decade, the various genomes sequencing projects fed the biological databanks with an extraordinary amount of data that remains of little use if not transformed into knowledge. Currently, the laborious process of annotation is carried out jointly by human experts and data-processing programs. Similarly, new technologies (proteomics, transcriptomics) start to produce mountains of data. The goal, from now, is more to track the activity of whole genomes, temporally and spatially than to thoroughly study biological objects taken separately. Knowledge is deduced from overall gene expression measurements in particular experimental contexts. The assumption is that a set of gene products is probably involved in a functional module when their levels of expression vary in a coordinated manner (Segal *et al.,* 2003). Work thus consists in two distinct phases: identifying these modules and then understanding their roles.

The first phase is now abundantly studied (Quackenbush, 2002). Numerous approaches dedicated to the acquisition, normalization, filtering and clustering of such high throughput results are available (Chuaqui *et al.,* 2002). In the end, treated data are more reliable and organized, but still very numerous. There is more than ever a need for automatic or semi-automatic approaches relying on structured and controlled vocabularies (ontologies) to analyze large quantities of data in order to discover meaningful patterns and rules (Attwood and Miller, 2001).

The THEA project is dedicated to the elaboration of tools and methods suited for the analysis of post-genomic data. In this paper, we present the first module developed in the frame of the project. It belongs to the field of knowledge discovery and is focused on the exploration and annotation of data generated by microarray experiments (Schena *et al.,* 1995).

## SYSTEMS AND METHODS

Two basic requirements of knowledge discovery are the access to the most complete and up to date information and its rapid availability (Fayyad *et al.,* 1996). In THEA, these requirements have led to the elaboration of ALLONTO, a dedicated data warehouse which stores selected data extracted from electronic resources, supplemented by a mediator which dynamically queries required and specific complementary informations over the internet.

In order to fully exploit data, knowledge discovery systems rely on a formal representation of information based on a well defined semantic (Simoff and Maher, 1998). This formal system is represented in ALLONTO by ontologies, which constitute a popular way to modelize biological concepts and their relationships.

### Ontologies of biological concepts

THEA is designed to make use of Ontologies described as Directed Acyclic Graphs (DAGs). A DAG is a structure composed of nodes (representing terms) and oriented arcs (representing relationships between terms) containing no cycle. This means that if there is a path from one node to another, then there is no way back. Such a modelization is very popular because it is intuitive, easily editable and less limited than hierarchical structures since terms can be source and target of many relationships. DAG based ontologies cover many biological domains (see for example the list collected by OBO at http://obo.sourceforge.net/list.shtml).

Presently ALLONTO includes two ontologies. Gene Ontology (GO) (Ashburner *et al.*, 2000) is a controlled vocabulary developed by a consortium of scientists. It can be used to describe ('annotate') a gene product in regard to its molecular functions (GO:MF), cellular localizations (GO:CL) and biological processes (GO:BP). Specific vocabularies dedicated to *Drosophila melanogaster* are developed by FlyBase (The_Flybase_Consortium, 2003), they describe the developmental stages and the anatomy of the fly: Drosophila Developmental Stages (FB:DDS) and Drosophila Gross Anatomy (FB:DGA), respectively. Progresses are being made to incorporate other ontologies as they develop.

Our database schema is designed as an extension of the GO one. This compatibility allows us to directly import the gene ontology from GO database as SQL tables (downloadable from http://www.godatabase.org/dev/database/archive/). FlyBase ontologies are reconstructed from flat files (available at http://flybase.bio.indiana.edu/docs/flydocs/flybase/controlled-vocabularies.txt) and integrated into the SQL tables.

**Ontologies associations**

Ontology constitutes a mechanism for expressing and sharing biological concepts which, in order to be useful, must be used as qualifiers for underlying data. Associations between gene products and GO terms are imported from text files elaborated by a growing number of biological databases (http://www.geneontology.org/doc/GO.current.annotations.shtml). Concerning *Drosophila* ontologies, associations are queried from the "Gene Expression" page of FlyBase (http://flybase.bio.indiana.edu/cgi-bin/expat).

**Genome data**

As no single source contains all the necessary information, one of the most fastidious tasks in functional genomics is finding the correspondences among the multiple identifiers of genes or gene products. To assist the knowledge discovery process, we have collected cross-links about all known genes, transcripts and proteins for nine organisms (*Homo sapiens*, *Mus musculus*, *Rattus norvegicus*, *Danio rerio*, *Fugu rubripes*, *Anopheles gambiae*, *D. melanogaster*, *Caenorabditis elegans* and *Caenorabditis briggsae*), from the Ensembl database (Hubbard *et al.*, 2002). Thereafter ALLONTO provides relationships between various identifiers from diverse sources, including organism-specific databases [FlyBase (The_Flybase_Consortium, 2003) for *D. melanogaster*, MGD (Blake *et al.*, 2003) for *M. musculus*, SGD (Weng *et al.*, 2003) for *Saccharomyces cerevisiae*, RGD (Steen *et al.*, 1999) for *R. norvegicus* and Wormbase (Harris *et al.*, 2003) for *C. elegans*] and general databases (e.g. SwissProt, PDB, PIR, PFAM, Ensembl, LocusLink, OMIM and EMBL).

GenBank files (Benson *et al.*, 2003), which contain annotations for raw genomic sequences, are also used to retrieve knowledge at both DNA and protein levels. Information concerning open reading frames (ORFs) include raw DNA sequences and physical maps. For corresponding proteins, translated sequences and gene symbols are collected. When available, references to Affymetrix probes identifiers are also extracted and stored. Specifically for *D. melanogaster*, we extract the identifiers of the Drosophila Gene Collection (DGC) cDNA clones (releases 1 to 3) (http://www.fruitfly.org/DGC) and link them to the corresponding gene products. Concerning *S. cerevisiae*, we extract the cross-links information, ORF name and localization from a flat file available at SGD

(ftp://ftp.yeastgenome.org/yeast/data_download/chromosomal_feature/chromosomal_feature.tab).

Data stored in our database contains only information needed by the data mining process. Other information, usually targeted for a human user are displayed when required as web pages retrieved by the mediator.

Regular import of external data is necessary to allow up-to-date analyses. However, drastic changes in data stored in public databases structures may occur. For example, in the Release 3 of FlyBase (March 2003), numerous genes have been merged and/or split, resulting in changes in >40% of the predicted proteins (Misra *et al.,* 2002). This can lead to important difficulties for the analysis of older experiments with a large majority of identifiers now obsolete. One then needs to track annotation history by collecting old or obsolete identifiers. We achieve this goal for *D. melanogaster*, by parsing a textual dump of the database (available at ftp://flybase.net/flybase/genes/genes.txt) which contains, for each entry, the corresponding FlyBase Id (identified by the special tag '*z') and the list of obsolete Ids (identified by the tags '*y'). This information, stored into ALLONTO, allows users to refer a gene product with any of its known identifiers.

**Classifications**

Basically, THEA takes the result of a hierarchical clustering performed on the data as primary input, in Newick format http://evolution.genetics.washington.edu/phylip/newicktree.html) or standard output of common classification programs like SOTA (Herrero *et al.,* 2001) or Cluster (Eisen *et al.,* 1998). It is also possible to analyze non-hierarchical clusters by building a fictive tree grouping the different classes. Alternatively, one can use the facilities proposed on the GEPAS site (http://gepas.bioinfo.cnio.es/) to build a hierarchy on top of classes generated by Self-Organizing Map (SOM) classification (Kohonen, 2001).

**ALGORITHM**

The Graphical User Interface of THEA allows users to explore biological data in a convenient way. It is possible to browse the ontologies, look for a particular field of knowledge and visualize associated leaves in the classification tree with colors markers. Successively Using a few terms will immediately reveal which of the clusters simultaneously pertain to different fields of knowledge, or if the classification can be broadly divided in different parts. This manual exploration is skewed as it is driven by the user's knowledge and his field of interest. To overcome this bias, THEA includes several data mining algorithms allowing an entire classification to be automatically annotated.

**Ontology-based cluster annotation**

For each major branch of the ontologies, choosing a unique label, for a cluster of co-expressed genes is difficult and it is illusory to try to elaborate a general and automatic tool to perform this task, as pertinence of the information is very user-dependent. It is one of the objectives of the project to take into account a user profile in the determination of labels. The current version does not include this feature and thus does not force the labeling of the clusters. Instead, a number of options allow to parameter the process. Two annotation processes have been found to be pertinent:

*Identification of common biological features*

For each cluster, THEA is able to extract the biological terms characterizing a given proportion of genes. The proportion is computed for each term and each cluster by dividing the number of genes associated to this term or one of its children by the total number of genes in the cluster. With a proportion of 1.00, a term will be used to label a cluster only if each of its members has the relevant association. This condition, which is very stringent, produces an annotation for a limited number of clusters only; lowering the proportion cutoff increases the number of named clusters but lowers the pertinence of these labels. This method has a tendency to name clusters with very generic terms. However, in some cases, using a high proportion cutoff allows the user to simplify the classification by labeling a cluster of genes with a unique term.

*Identification of under and over represented terms*

For each cluster of transcripts, THEA performs a statistical analysis under the null hypothesis of a uniform distribution of annotations. A given cluster is considered significantly enriched for a term if the number of transcripts associated with it exceeds the number expected by chance. The hypergeometric distribution should be used in the calculation but, this law is indeed slow to calculate when one has to handle several thousand clusters. The binomial law, which is less computing-intensive and constitutes a good approximation of the hypergeometric law when the population is large, can most of the time, be used instead. THEA offers the possibility to choose between these two statistical laws. It has to be stated that all P-values given in this paper were computed with the binomial law. For a given term present in the total population with a frequency $p$, the laws calculate the probability (or *P-value*) of observing by chance at least $k$ transcripts annotated with this term in a cluster of size $n$. Enriched terms are extracted by selecting those associated with a *P-value* under a given (adjustable) cutoff. The same principle is used to highlight under-represented terms by calculating the probability of observing a maximum of $k$ transcripts annotated with a term in a cluster. It should be noted that absent calls or unreliable measurements (e.g. due to bad reporters) can make some transcripts show undetectable expression levels resulting into under-evaluation of the number of represented terms. This bias may increase the confidence level of the over-represented genes but lead to unreliable conclusions about under-represented terms, unless the hypothesis of uniform distribution of annotations is restricted to detected transcripts (see below).

The null hypothesis can be relative either to all genes associated to the ontologies, to the list of genes included in the classification or to a specific user-defined list. Using all GO annotations is a method commonly used (Draghici *et al.,* 2003) but can be skewed if the assayed genes represent only a subset of the genome. For example, if using an array bearing reporters for each transcript known to or thought to be regulated during mitosis, most of the clusters obtained would show an artefactual over-representation of the term "mitosis" whatever the classification method used. In this case, it is better to perform the calculation under the null hypothesis of a uniform distribution of the associations restricted to the transcripts assayed by the array. Another option is to use as reference group, the set of transcripts included in the classification. Such an analysis does not give an idea of the biological concepts of importance in the experimental conditions studied but allows the user to automatically detect where the related transcripts are located in the classification and thus how they behave transcriptionaly. THEA allows the user to choose either solution.

**Analysis of transcription maps**

A second type of research handled by THEA consists of highlighting the possible correlations between expression levels of certain genes and their localization on the genome. A technique similar to the one used for the identification of an over-representation of terms in a cluster is used. Under a null hypothesis of a uniform distribution of genes on each chromosome, we can calculate a *P-value* for a group of co-localized genes to appear in the same cluster. The user has first to specify the maximum distance $d$ between two genes to consider them co-localized. The measure of distance that we have chosen to use is the number of intergenic regions separating two genes: if one attributes an index for each gene on a chromosome (the closest gene to the 5'-end is #1, the second closest #2, etc), then the distance between two genes is the difference of their indices. For a given gene and a given distance $d$, there exist, in the whole genome, a maximum of *2d* other genes in a distance less or equals to $d$ (these neighborhood genes are those which are considered as co-localized). For each gene in a cluster we determine the number $c$ of co-localized genes present in the same cluster and we use the binomial law to compute the probability of observing at least $c$ genes in a cluster of size ($n$-1) by chance, considering a frequency of 2d/G (G representing the number of genes in the genome). This procedure can only detect groups of at least three genes (the gene taken as reference plus two genes present in its neighborhood). Genes that are considered co-localized under a certain cutoff are highlighted with a marker similar to that used to label genes on the classification (Fig. 1).

**IMPLEMENTATION**

THEA is based on DAG-Edit, an open source software available from sourceforge (http://sourceforge.net/projects/geneontology/). It makes use of several other open source software including IzPack (graphical installer), Xerces (XML parser), Regexp (regular expression package), Ant (XML based makefile), Batik (SVG toolik), Connector/J (java driver for MySQL) and colt (statistical package). The relational database MySQL is used to store the information needed by the software.

The architecture of THEA is based on the Model-View-Controller (MVC) design paradigm. Its interface is composed of plugins that users can choose to activate or not. Each plugin is dedicated to the display and/or edition of various information concerning both the underlying biological objects (genome data, ontologies and clustering results) or the software itself (captions, options, search parameters, etc).

Figure 1 shows a typical layout of THEA's user interface with six activated plugins: Terms Viewer (display of loaded ontologies), Gene Displayer (view of associated gene products), Tree View (visualization of clustering results), Classification Caption (information on the classification), Search Panel and DAG Viewer (a trimmed ontology leading to the selected term). From this integrated environment, users can explore the data in multiple ways, including queries of gene products associated with one or several terms, localization of gene products in the classification, visualization of terms characterizing gene products, automatic labeling of groups of co-expressed genes, simplification of classification tree by hiding details for clusters of similar genes, etc.

**BIOLOGICAL APPLICATION**

A dataset consisting of 34 Affymetrix Genechip Drosophila Genome arrays hybridized with a time series of RNA extracts from flies challenged with bacterial or fungal infection (De Gregorio *et al.*, 2001) has been chosen to demonstrate the functionalities of THEA

**Annotation comparison**

The reanalysis of these data lead to the selection of 1270 regulated genes (see supplementary materials). Annotating this set with THEA takes ~1 mn on an standard computer and produces the following results: 64.4 % of the genes are associated to at least one term of GO, 13.9 % to at least one term of the FlyBase ontologies (FB) and 13.5 % to both. De Gregorio *et al.* had selected a set of 400 over- or under-expressed genes (List0). In their table 1 (p 12592) they manually identified groups among these genes, using the appropriate GO terms when available. From this table, we selected three groups for demonstration purposes (List1), comprising 7 peptidoglycan recognition proteins (PGRPs, GO:0016019), 8 Serine-protease inhibitors (serpins, GO:0004868) and 15 Antimicrobial peptides (AMPs, GO:0003795). These GO terms were used to extract the corresponding list of associated genes from our reanalyzed data according to THEA (List2). Merging the two lists resulted in a set of 50 genes. Figure 2 shows their expression profiles and color-coded GO annotations: PGRPs (green), serpins (blue) or AMPs (pink). Genes from List1 are depicted in pale tones, while genes found in List2 are highlighted with dark tones. Genes from List0 are labeled with grey bars.

The first and most obvious observation is that all the genes labeled in pale tones (original paper) are also labeled in dark tones (THEA) , with the exception of *CG10812*. This shows that what had been done painfully by hand can be retrieved with THEA. *CG10812* is a counter-example: in the original paper it was included in the AMPs (GO:0003795), when in fact its GO annotations are "defense response" (GO:0006952) and "defense/immunity protein activity" (GO:0003793), this last term being the parent of AMP. Therefore, the association of *CG10812* to the term "antimicrobial peptide activity" (GO:0003795) was incorrect in GO sense, but might have been manually added by the authors according to their own expertise.

The second interesting observation is that seven genes are labelled grey and dark but not pale: *PGRP-SC1b, CG12780, CG13422, TepII, TepIV, CG16756*, and *nec*. These genes were thus not associated in the original paper to any of the three selected GO terms. Such discrepancy might result from an improvement in GO since the original analysis, from errors in the ontology (manually corrected by the authors), or simply from overlooking of such associations (a frequent problem with manual annotations). For example, *PGRP-SC1b* shows similarity of sequence with other PGRP coding genes (Werner *et al.,* 2003); this was probably overlooked by the authors, but is automatically retrieved by THEA. *CG12780* and *CG13422* had been included in a non-GO class ("GNBP" for Gram-Negative Binding Proteins), corresponding to the GO term "Gram-negative binding activity" (GO:0008368"), to which they are presently associated. They were retrieved by THEA because GO:0008368 is a child of AMP. The authors described GNPB as involved in recognition and phagocytosis rather than in direct antimicrobial activities. *TepII* and *IV* had also been included in a non-GO class ("complement-like"), while they are presently described as being AMPs (Lagueux, Perrodou *et al.,* 2000). Finally, the authors mentioned that *nec* codes for a serpin, but they included it under the Toll-Pathway term (GO:0008063), an association that THEA is also able to detect (data not shown).

The last category of genes present in Figure 2 is labeled only with dark tones (they had not been retained in the original paper: no grey tab). In fact, the authors selected 400 genes among more than 13000 represented on the chip, which was already a considerable amount for a hand annotation process. However, since THEA allows immediate annotation of large datasets, we were not concerned about drastic limitation of the number of immune-responsive genes to be retained and therefore we analyzed a group three times larger the one described in the original publication.

In conclusion, this comparison shows that THEA can emulate quite well what can be done by a human expert. If anything it is more complete and remarkably faster and much easier.

**Searching for immunity-related genes**

Using THEA's Search Panel or Terms Displayer allows the selection of a term of interest and the labeling of the associated genes in the classification. In this example, the selection of "immune response" (GO:0006955) labels 42 gene products, unevenly distributed in the classification (red tabs in Figures 1 and 3). Indeed the majority of them (35) is found in the lower part of the tree, which groups genes globally overexpressed after an immune challenge. Closer inspection reveals that most (25) are further regrouped in two clusters of closely co-regulated genes together with genes not known to participate in the immune response (see blue box in Fig. 3).

Overall, THEA allows to rapidly pick out genes of interest among the whole clustering, which helps to get a general idea of their behavior. In our example, we discovered in a matter of minutes that, among regulated genes, those known to participate to the immune response are (1) as expected, well represented (42 out of 86 in the genome), (2) only weak minority (42 out of 1270 regulated), (3) mostly found among the genes overexpressed (35 out of 42) and (4) regrouped in two closely associated clusters (25 out of 42). Globally, this implies that our current GO-documented knowledge of the immune response is only able to account for a small fraction of the variability observed.

**High-level interpretation**

While the previous feature is useful when one knows what to look for, such an approach is still obviously skewed. To avoid this effect we can ask THEA to label the nodes of the classification, with over and under-represented terms. Figure 3 presents a fully labeled version of the classification. Overall 30 (61.2 %) of these nodes are labeled, with a mean number of 6.2 terms per node (max:9). Most of these terms come from GO (121 labels): 48 from GO:MF, 37 from GO:CC and 36 from GO:BF. Of the 37 FB names, 27 come from FB:DGA and 10 from FB:DDS (see supplementary table S1). It might be noted that the six nodes named with a FB:DDS ontology under-represented term are occurrences of a single term: 'embryonic stage'. This observation is coherent with the fact that the original study was conducted with RNAs extracted from adult flies.

This allows a general view of the behavior of the transcriptome in the experimental conditions, by thinking in terms of concepts (the ontology terms displayed) rather than in terms of genes. A striking observation is that, when going from the root to the genes, the terms used to describe the successive nodes often become more precise. Therefore, as the clusters of genes have more coherent expression profiles, the precision of the associated information grows.

Terms at the root show that, in *Drosophila*, genes regulated after infection are significantly enriched in genes coding for enzymes (GO:0003824), for proteins located in the mitochondrial inner membrane (GO:0005743), involved in the defense response (GO:0006952) and expressed in the fat body (FBbt:00005066) ; while those coding for proteins having nucleic acid binding activities (GO:0003676), located in the nucleus (GO:0005634), involved in cell proliferation (GO:0008283) and expressed during the development (GO:0007275) are under-represented (see labels in Fig. 3). Taken together, this shows that the immune response induces major modification in the enzymatic capacities of the fly, especially in the mitochondria, while proteins located in the nucleus and involved in nucleic-acid-related processes are rather unaffected. It is also observable that, not surprisingly, infection induces a modification of the levels of expression of defense response genes and that it highly affects the fat body's transcriptome, a fact that is well documented (Hoffmann, 2003).

A general observation is that under-represented terms are only found in the upper branch of the tree, which corresponds to genes globally repressed after infection (see heatmap). The correlation between under-expression and under-labeling is not easy to interpret, but biological meanings may exist. Labels used to describe this upper branch indicate that infection induces the repression of energetic pathways, a fact that was mentioned by De Gregorio *et al*. (2003) .They interpreted it as resulting from turning down unessential metabolic pathways in order to redirect the energetic and tranductionnal capacities towards defense response. Alternatively, we can postulate that infected flies might be a little cataleptic and/or feed less than healthy ones, thus needing less metabolic capabilities. The fact that genes expressed in the indirect flight muscles are also found to be repressed (an observation not made by the authors) can be interpreted in a similar manner: sick flies would be less mobile; moreover they might need to reallocate the huge amounts of amino acids stored in their muscles to synthesize defense response proteins in vast quantities. Induced genes are significantly enriched in genes coding for antibacterial peptides, involved in immune response and expressed in the fat body (Fig. 3, lower part). This perfectly reflects what is known of the humoral immune response in *Drosophila*: pathogens induce fast and massive release of small peptides having antimicrobial properties (AMPs) in the circulatory system of the fly (the hemolymph), the fat body being the major site of this AMP synthesis and secretion (Hoffmann, 2003). An interesting observation is that this branch is also labeled with the term "translocon". Therefore THEA automatically detects that the translocon components and AMP coding genes are upregulated in a similar manner, a fact that was not reported in the original publication. The translocon is a multiproteic complex involved in the early steps of the secretion of proteins in the extracellular medium (Meacock *et al.,* 2000). We interpret this observation as a functional adaptation of the flies to infection: the synthesis of huge amounts of circulating proteins as AMPs is likely to require a dramatic increase in the secretion capacity of the flies, hence the increase in translocon quantities. It is tempting to propose that this increased secretion capacity is mainly located in the fat body.

**CONCLUSION**

The use of automatic description of the nodes to interpret massive data-sets in terms of concepts, only briefly described here, changes drastically our way of analyzing chip data. The use of statistical criteria rather than manual inspection according to selected hypotheses makes this analysis as unbiased as possible. Using a previously published dataset we were able to reach similar conclusions and detect well-known characteristics of the immune response (as the overexpression of numerous AMPs) as well as less obvious observations (as the upregulation of genes participating in the secretion pathway), which were not reported in the original publication. The only limitation is inherent to the nature of the analysis: it cannot be better than the ontologies used, but ontologies are improving everyday.

## ACKNOWLEDGMENTS

This work is supported by the French Bioinformatic Program of the Ministry of Education 2002-2004.

**Fig. 1: General view of THEA's user interface, displaying fully annotated classification of the 1270 regulated genes.**

This example configuration of THEA's user interface shows six activated plugins. Some of them are inherited from DAG-Edit (http://sourceforge.net/project/showfiles.php?group_id=36855): The **Terms viewer** (top left), allows to graphically browse through the loaded ontologies and select individual terms. The **DAG Viewer** (top right), displays a trimmed ontology leading to the selected term.

Other plugins have been developed as part of the THEA project: The **Gene Displayer** (middle left) displays the genes associated to a given term and, eventually, its children (here the 86 genes associated to the term 'immune response' and their children are displayed) and provides access to the relevant publications and databases *via* Internet. The **Search Panel** (middle top) allows to search for specific terms or character strings among genes or terms. Complex queries are also possible. Here the term "immune response" has been selected. The **Classification Tree Viewer** (bottom right) allows the display of hierarchical clustering results of gene expression profiles. It is possible to navigate in the tree by zooming on the nodes. The expression profiles are displayed on the right in the classical form of a heatmap (*à la* TreeView (Eisen *et al.* 1998)). Genes associated to the selected term and their children are labeled with colored boxes. As in the Gene Displayer, right clicking on the genes gives access to any relevant information. The names automatically assigned to each cluster are displayed in this plugin as well. The **Classification Caption** (bottom left) displays the legends of the color-coded labels as well as some statistical information.

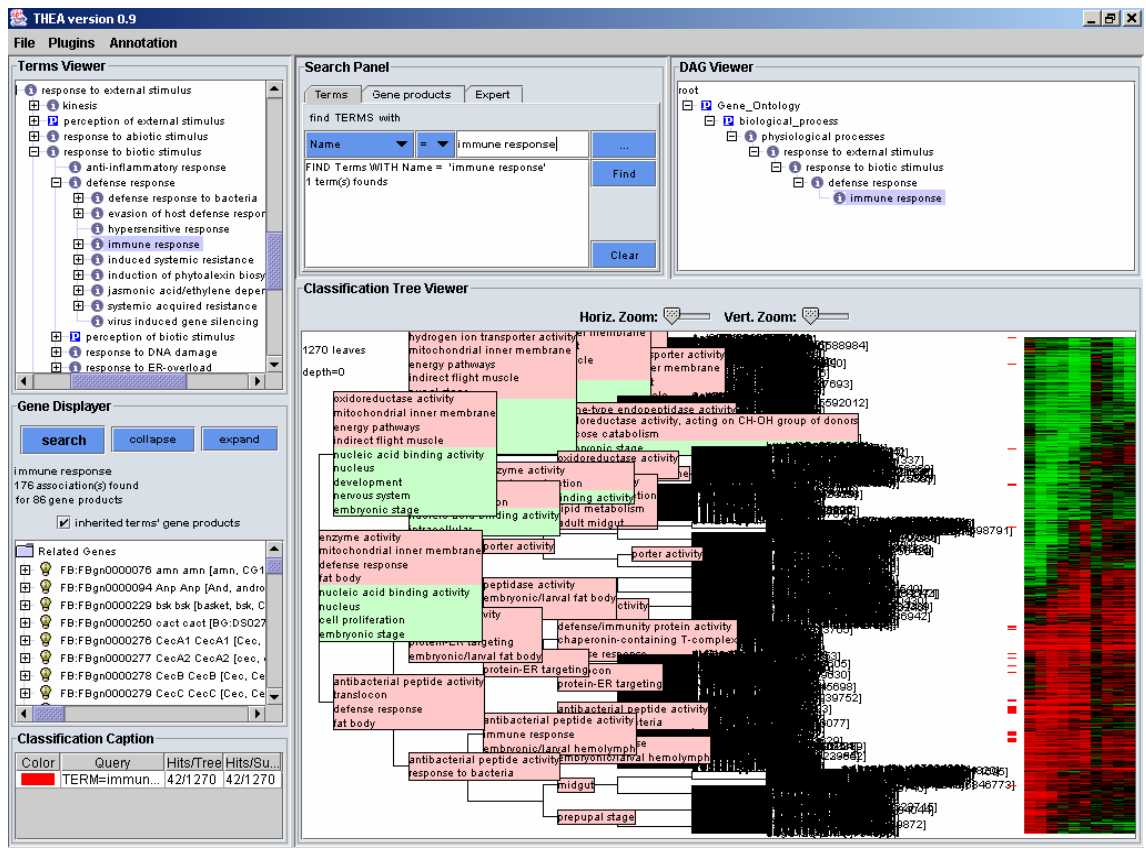

**Fig. 2: Annotation of a subset of genes with GO, comparison of the analysis taken from the original publication with THEA (see text for details).**

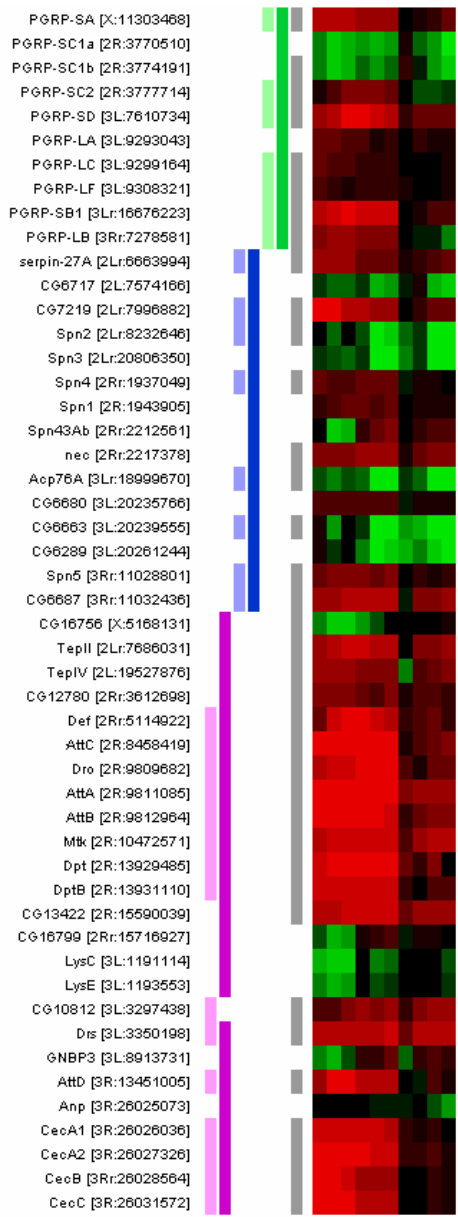

**Fig. 3: Annotated classification of the 1270 regulated genes as displayed in the Classification Tree Viewer.**

Genes associated to the term "immune response", selected *via* the Search Panel, are automatically labeled (red labels on the right of the classification), multiple selections being allowed (Fig. 2). For each node, terms statistically over-represented are shown in pink boxes while terms under-represented are in green boxes (see text for further details).

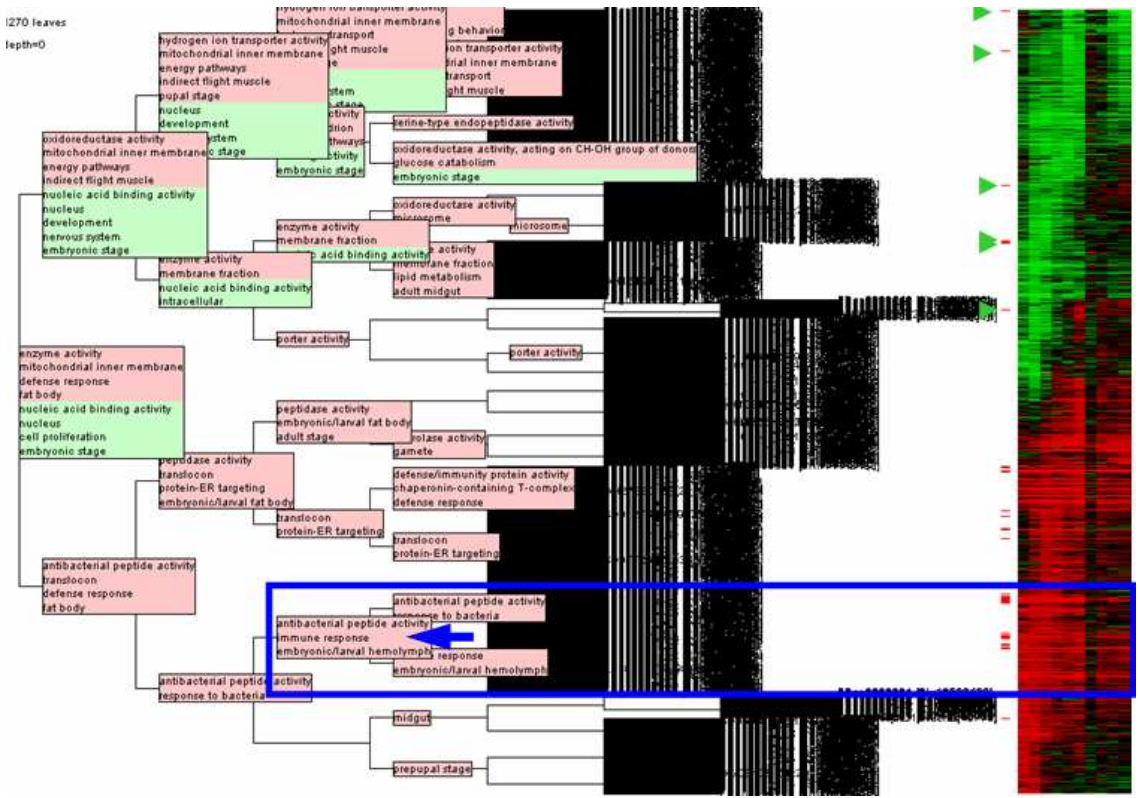